\documentclass[aps,physrev,twocolumn,groupedaddress]{revtex4-2}

\usepackage{graphicx}
\usepackage{dcolumn}
\usepackage{bm}
\usepackage[version=3]{mhchem} 
\usepackage{makecell}
\usepackage{xcolor}
\usepackage{soul}
\usepackage{cleveref}
\usepackage{subcaption} 
\usepackage{multirow}
\usepackage{float}

\begin{document}

\newcommand{\rcom}[1]{\textcolor{red}{#1}}

\title{Bond strengths in solids computed from a Wannier-type construction of local vibrational modes}

\author{Mateusz Mojsak}
\affiliation{School of Chemistry, University of Birmingham, Birmingham B15 2TT, United Kingdom}

\author{Elfi Kraka}
\affiliation{Department of Chemistry, Southern Methodist University, 3215 Daniel Ave, Dallas, Texas 75275-0314, United States}

\author{Adam A.\ L.\ Michalchuk}
\email[Contact author: ]{a.a.l.michalchuk@bham.ac.uk}
\affiliation{School of Chemistry, University of Birmingham, Birmingham B15 2TT, United Kingdom}
\affiliation{Federal Institute for Materials Research and Testing (BAM), 12489 Berlin, Germany}

\date{\today}

\begin{abstract}
We introduce a Wannier-type formulation of periodic local vibrational mode theory that yields real-space-localized vibrational modes associated with individual internal coordinates in crystalline solids. These modes are constructed as locally coherent superpositions of wavevector-resolved local modes, yielding a smooth and gauge-consistent real-space representation without the need for additional phase-fixing procedures. The resulting Wannier-type local modes provide well-defined force constants and frequencies that enable robust, chemically interpretable measures of bond and interaction strengths in periodic systems. Moreover, our framework demonstrates that phonon dispersion behavior makes important contributions to the bond and interaction strengths calculated \textit{via} local vibrational mode theory.  We demonstrate the method for representative ionic and covalent systems, including \ce{MgO}, tetrahedrally-coordinated \ce{C}, \ce{Si}, \ce{SiC}, and two polymorphs of \ce{CaCO3}. Our framework establishes a direct analog of molecular local modes for fully periodic systems and opens new avenues for quantitative bonding analysis in crystalline materials.
\end{abstract}

\maketitle

\section{Introduction}{\label{sec:intro}}

Solid state materials underpin much of modern technology, from new energy harvesting and conversion devices to electronic and catalytic materials. The diverse applications of solids arise from the subtle ways in which variations in interatomic interactions influence structure, stability, and macroscopic functional properties. Consequently, understanding and quantifying the nature of specific interatomic interactions and how they give rise to functional behavior is essential for the discovery and development of new materials with enhanced or tunable properties.\cite{wuttig_AdvMater_2023,wuttig_AdvMater_2022,raty_AdvMater_2019}

The contributions of specific interactions to functional properties are often closely linked to the strength of the interaction. This relation has, for example, inspired recent efforts to machine-learn functional properties using descriptors derived from interatomic interaction strengths.\cite{naik_arxiv_2026} However, despite the apparent conceptual simplicity of interatomic interaction strengths, quantifying them in crystalline solids remains a fundamental challenge. In a periodic system, interatomic interactions are embedded within a many-body environment, and no unique observable can be assigned to a single bond or interaction without invoking a specific partitioning of, e.g., real space,\cite{wilson_cgd_2024} charge density,\cite{1998Bader,johnson_revealing_2010} or degrees of freedom.\cite{hempelmann_advmater_2021} Perturbing or “breaking” a local interaction in a solid necessarily induces structural and electronic relaxation across the surrounding structure, thereby blurring the distinction between local chemical effects and the collective elastic response. As a result, any measure of bond strengths in a periodic system is inherently method-dependent, and different approaches often emphasize complementary but non-equivalent physical aspects of the same interaction.

A wide range of methods have been developed to characterize bonding in periodic systems. The majority of approaches are based on the analysis of electronic structure quantities obtained from static crystal structures, including topological descriptors, e.g., quantum theory of atoms in molecules (QTAIM),\cite{Bader1994} electron localization function (ELF) analysis,\cite{becke_simple_1990} non-covalent interaction (NCI) descriptors,\cite{johnson_revealing_2010,saleh_revealing_2012} and orbital-based approaches\cite{muller_chemsci_2025} (e.g., natural bonding orbital\cite{dunnington_jctc_2012} or crystal orbital Hamilton population (COHP) analyses\cite{maintz_JCC_2016}). These methods have seen widespread success for obtaining chemically-informed insight into bonding patterns and interaction types, though they typically infer interaction strength through indirect (and often non-intuitive) metrics.  

Local vibrational mode theory (LVMT) offers a powerful approach to characterize bond strengths based on the vibrational (normal-mode) behavior of the system, yielding local stretching force constants that serve as intuitive and quantitative measures of intrinsic bond strength.\cite{kraka_wires_2020} This approach has been widely used for studying molecules in isolation, including for the study of luminescence behavior in lanthanide complexes,\cite{saraiva_role_2025} for interpreting intramolecular vibrational relaxation pathways,\cite{mallon_anharmonically_2025} for re-assessing $\mathrm{p}K_\mathrm{a}$ descriptors in small molecules,\cite{quintano_pka_2023} and for analyzing the strengths of explosophoric bonds in high energy-density molecules.\cite{christopher_frontiers_2021} More recently, LVMT was extended to investigate bonding in the solid state, based on the analysis of normal mode eigenvectors at the Brillouin zone center. This extension enabled the analysis of diverse systems including the effect of crystallization on covalent bonding in the uranyl ion (\ce{UO2^2+})\cite{bodo_jcompchem_2024} and the influence of pressure on \ce{I-O} bonding in \ce{Mg(IO3)2},\cite{bodo_prb_2026} demonstrating the significant potential of LVMT as an easy-to-interpret descriptor for studying the role of specific interactions in a variety of chemical and physical problems.

\begin{figure}
    \centering
    \includegraphics[width=0.7\linewidth]{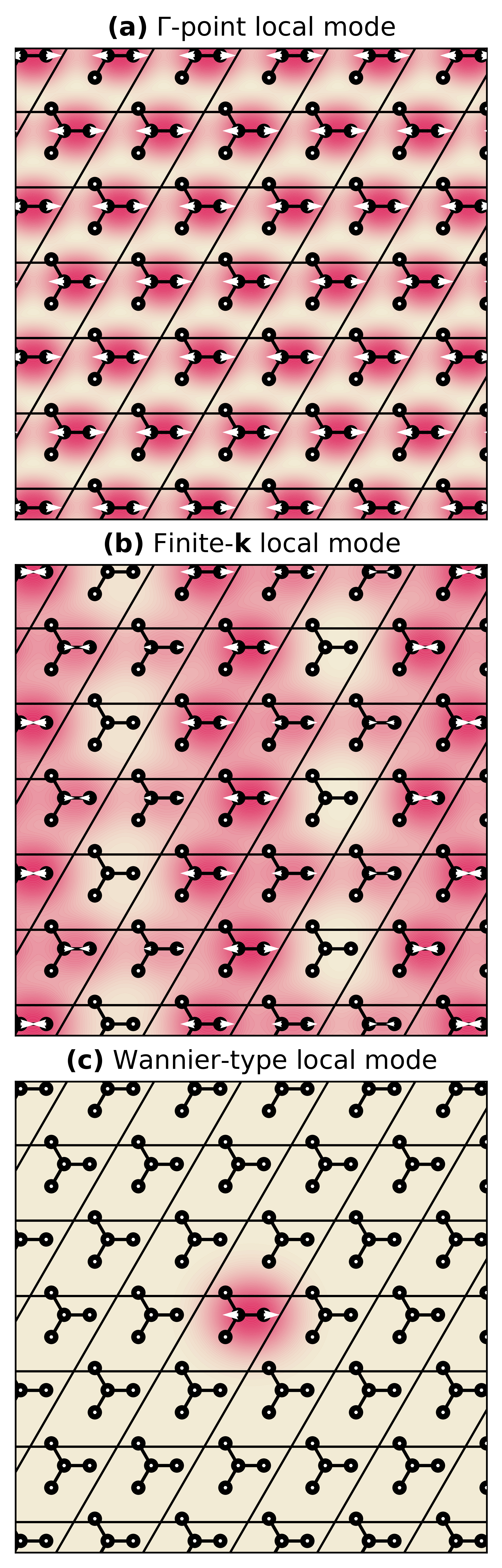}
    \begin{subfigure}{0pt}
        \phantomcaption
        \label{fig:lmodes:gamma}
    \end{subfigure}
    \begin{subfigure}{0pt}
        \phantomcaption
        \label{fig:lmodes:finitek}
    \end{subfigure}
    \begin{subfigure}{0pt}
        \phantomcaption
        \label{fig:lmodes:wannier}
    \end{subfigure}
    \caption{Illustration of the displacement patterns associated with wavevector-resolved local modes---(a)~$\Gamma$-point and (b)~finite-wavevector---and (c)~the~Wannier-type local modes introduced in this work. Atomic displacements (white arrows) are repeated periodically throughout the entire lattice for wavevector-resolved local modes, but are localized to a single unit cell for the Wannier-type mode. The background color represents the local displacement magnitudes.}
    \label{fig:lmodes}
\end{figure}

Reflecting the wavevector-dependence of normal vibrational modes in periodic systems, we have extended LVMT by defining wavevector-resolved local modes for arbitrary k-points in the Brillouin zone.\cite{mojsak_jctc_2026} Using this formulation, one can track how the internal-coordinate character of normal modes evolves across the Brillouin zone, probe the chemical origins of phonon dispersion behavior, and hence explicitly and quantitatively study how local structural distortions contribute to collective vibrational phenomena in periodic solids. However, while these wavevector‑resolved local modes are localized to specific internal coordinates within each unit cell, they remain delocalized across the periodic lattice, \Cref{fig:lmodes:gamma,fig:lmodes:finitek}. Consequently, their associated force constants correspond to collective rather than individual bond distortions. This naturally raises the question as to whether, and how, one can recover real‑space localized vibrational descriptors that accurately reflect individual bond or interaction strengths in periodic systems that allow for a one‑to‑one comparison with local mode force constants in isolated molecular systems.

As the next step in our development of LVMT for periodic systems, we introduce here a Wannier-type formulation of periodic LVMT that produces real-space localized vibrations associated with individual internal coordinates centered on a single unit cell, \Cref{fig:lmodes:wannier}. Analogous to electronic and phonon Wannier functions, these modes are obtained by forming locally coherent superpositions of the wavevector-dependent local modes over the first Brillouin zone. The phases of the constituent modes are fixed relative to a chosen reference unit cell, producing a smooth, gauge-consistent representation across k-points without the need for additional gauge-fixing procedures. This procedure yields displacement patterns that are spatially concentrated around a single unit cell, affording physically meaningful local mode force constants that provide a direct metric of bond strength in solids in intuitive units of elastic stiffness. This approach naturally respects both the local character of chemical interactions and the translational symmetry of the crystal, providing a robust and chemically intuitive framework for analyzing bond and interaction strengths in periodic systems.

\section{Theoretical Background}

\subsection{Wavevector-resolved local vibrational modes}
\label{sec:wvresmodes}

\subsubsection{Vibrational dynamics expressed in internal coordinates}

Wavevector-resolved local vibrational modes are obtained from the eigenvectors and eigenvalues of the reciprocal-space force constant matrix expressed in Cartesian coordinates, $\mathbf{f}^x(\mathbf{k})$.\cite{mojsak_jctc_2026} For a given wavevector $\mathbf{k}$, the elements of $\mathbf{f}^x(\mathbf{k})$ are the Fourier transforms of the real-space force constants, obtained as the second derivatives of the total electronic energy $E$ with respect to atomic displacements $x_{\kappa \alpha}$,\cite{srivastavaPhysicsPhonons2019}
\begin{equation} \label{eqn:fx}
    f^x_{\kappa \alpha, \kappa'\alpha'}(\mathbf{k}) = \sum_{\mathbf{R}_\lambda} \frac{\partial^2 E}{\partial x^{\mathbf{0}}_{\kappa \alpha} \; \partial x^{\mathbf{R}_\lambda}_{\kappa'\alpha'}} \; e^{i\mathbf{k} \cdot \mathbf{R}_\lambda},
\end{equation}
where $\kappa$, $\kappa'$ label atoms in the unit cell and $\alpha$, $\alpha'$ denote Cartesian axes. $\mathbf{R}_\lambda$ are the Bravais lattice vectors. We define a reference unit cell at $\mathbf{R}_\lambda=\mathbf{0}$ with the $\lambda$-th periodic image at distance $|\mathbf{R}_\lambda|$. The Cartesian normal modes at wavevector $\mathbf{k}$ are obtained by solving the generalized eigenvalue problem
\begin{equation} \label{eqn:diag}
\mathbf{f}^x(\mathbf{k}) \; \mathbf{L}(\mathbf{k}) = \mathbf{M} \; \mathbf{L}(\mathbf{k}) \; \mathbf{K}(\mathbf{k}),
\end{equation}
where $\mathbf{M}$ is the diagonal matrix of atomic masses, the columns of $\mathbf{L}(\mathbf{k})$ are the eigenvectors, which describe the Cartesian atomic displacement patterns of the harmonic normal modes, and the diagonal elements of $\mathbf{K}(\mathbf{k})$ are the squared harmonic normal mode frequencies. In a normal coordinate basis, the matrix $\mathbf{K}(\mathbf{k})$ represents the force constant matrix of the harmonic normal modes,
\begin{equation}
    \mathbf{K}(\mathbf{k}) = \mathbf{L}^\dagger(\mathbf{k}) \; \mathbf{f}^x(\mathbf{k}) \; \mathbf{L}(\mathbf{k}),
\end{equation}
with its inverse representing the compliance matrix. 

It is worth noting that, when studying solid-state normal modes, one typically diagonalizes the dynamical matrix, 
\begin{equation}
    \mathbf{W}(\mathbf{k}) = \mathbf{M}^{-1/2} \; \mathbf{f}^x(\mathbf{k}) \;\mathbf{M}^{-1/2}
\end{equation}
to obtain normal mode frequencies and eigenvectors. This yields the same eigenvalues as \Cref{eqn:diag} but returns mass-weighted eigenvectors. Hence, by using \Cref{eqn:diag} we directly access a set of eigenvectors that correspond to the physical atomic displacements.

Using a $\mathbf{k}$-dependent equivalent of the Wilson $\mathbf{B}$ matrix\cite{wilsonMolecularVibrationsTheory1955} with elements for the $n$-th internal coordinate $q_n$ defined as\cite{mojsak_jctc_2026} 
\begin{equation} \label{eqn:bmat}
    B_{n,\kappa\alpha}(\mathbf{k}) = \sum_{\lambda}\frac{\partial q_n}{\partial x_{\lambda\kappa\alpha}} \; e^{i\mathbf{k} \cdot \mathbf{R}_\lambda},
\end{equation}
we can map the normal mode eigenvectors at wavevector $\mathbf{k}$ onto the internal coordinate representation,
\begin{equation} \label{eqn:dmat}
    \mathbf{D}(\mathbf{k}) = \mathbf{B}(\mathbf{k}) \; \mathbf{L}(\mathbf{k})
\end{equation}
where the row $\mathbf{d}_n(\mathbf{k})$ stores the contributions of internal coordinate $q_n$ to the normal modes.

\subsubsection{Wavevector-resolved local mode properties}

Following the adiabatically-relaxed local mode construction,\cite{konkoliNewWayAnalyzing1998-I,krakaLocalVibrationalMode2022} we can define the adiabatic atomic displacement vector for the $n$-th wavevector-resolved local mode as,
\begin{equation} \label{eqn:ax}
    \mathbf{a}^x_n(\mathbf{k}) = \mathbf{L}(\mathbf{k}) \cdot \frac{\mathbf{K}^{-1}(\mathbf{k}) \; \mathbf{d}_{n}^\dagger(\mathbf{k})}{\mathbf{d}_{n}(\mathbf{k}) \; \mathbf{K}^{-1}(\mathbf{k}) \; \mathbf{d}_{n}^\dagger(\mathbf{k})},
\end{equation}
with an associated force constant
\begin{equation} \label{eqn:k}
    k^a_n(\mathbf{k}) = \left[ \mathbf{d}_{n}(\mathbf{k}) \; \mathbf{K}^{-1}(\mathbf{k}) \; \mathbf{d}_{n}^\dagger (\mathbf{k}) \right]^{-1}
\end{equation}
and local mode frequency
\begin{align} \label{eqn:omega}
\omega_n^a(\mathbf{k}) 
  &= \sqrt{m_n^{-1} \, k_n^a(\mathbf{k})} \\
\text{where} \qquad 
m_n^{-1} 
  &= \mathbf{b}_n(\mathbf{k}) \, \mathbf{M}^{-1} \, \mathbf{b}_n^\dagger(\mathbf{k}).
\end{align}
The effective mass $m_n$ of the corresponding internal coordinate $q_n$ is obtained using the rows $\mathbf{b}_n(\mathbf{k})$ of the $\mathbf{B}(\mathbf{k})$ matrix, and acts to re-introduce the correct inertial weighting. Moreover, $m_n$ is $\mathbf{k}$-independent, see ESI~S1.

In a periodic system, the local mode vectors $\mathbf{a}^x(\mathbf{k})$ that describe the atomic displacement patterns within a reference unit cell also determine the corresponding displacements in all other unit cells \textit{via} the Bloch phase factor $e^{i\mathbf{k}\cdot\mathbf{R}}$. Consequently, for any atom $\kappa$, its real-space displacement in unit cell $\mathbf{R}_\lambda$ associated with a local mode $n$ at wavevector $\mathbf{k}$ is given by
\begin{equation}
\mathbf{u}_{n,\lambda\kappa}(\mathbf{R_\lambda};\mathbf{k}) = \mathbf{a}^x_{n,\lambda\kappa}(\mathbf{k}) \; e^{i\mathbf{k}\cdot\mathbf{R_\lambda}}.
\end{equation}
Hence, while wavevector-resolved local modes are localized on particular internal coordinates in each unit cell, they are necessarily delocalized across the entire periodic lattice, consistent with the underlying periodicity of the normal modes. 

\subsection{Real-space-localized internal coordinate vibrations in solids}

\subsubsection{Wannier-type construction of real-space-localized vibrational modes}
A periodic local vibrational mode can be further localised to an arbitrary unit cell at real space position $\mathbf{R}_\lambda$ by creating a coherent superposition of the wavevector-resolved local modes through an inverse Fourier transform. Correspondingly, we define a real-space local mode associated with internal coordinate $q_n$ analogously to a Wannier function as
\begin{equation}
\label{eqn:realspacelocalmode}
\mathbf{\tilde{a}}^x_n(\mathbf{R}_\lambda) = \frac{1}{\sqrt{N_k}} \sum_{\mathbf{k}} \mathbf{a}^x_n(\mathbf{k}) \; e^{-i\mathbf{k}\cdot\mathbf{R}_\lambda},
\end{equation}
where the sum runs over the $N_k$ wavevectors used to sample the Brillouin zone.\cite{wannier}

In our formulation of periodic local vibrational mode theory, the internal coordinates used to construct the local modes are defined in the immediate vicinity of a reference unit cell at $\mathbf{R}_\lambda=\mathbf{0}$.\cite{mojsak_jctc_2026} This definition offers a fortuitous restriction on the gauge freedom of \Cref{eqn:realspacelocalmode} by inherently fixing the phase origin of each local mode to the same reference cell. All contributions from periodic images therefore enter with consistently-defined Bloch phase factors and thus ensure that the $\mathbf{k}$-dependent local modes transform smoothly across the Brillouin zone. As a result, the wavevector-resolved local modes form a gauge-consistent representation, enabling their direct Fourier transformation into real space without the need for additional gauge‑fixing localization procedures. This property represents a key advantage of constructing Wannier-type localized vibrational modes from the chemically defined local modes rather than directly from the normal mode eigenvectors, which are only defined up to an arbitrary $\mathbf{k}$-dependent phase. 

The phase convention of the wavevector-resolved local modes renders our Wannier-type functions $\mathbf{\tilde{a}}^x(\mathbf{R}_\lambda)$ maximally coherent for the representations of the internal coordinates that are used to construct the $\mathbf{B}(\mathbf{k})$ matrix. 
By defining our internal coordinates as being centered on the reference cell at $\mathbf{R}_\lambda=\mathbf{0}$, \Cref{eqn:realspacelocalmode} for that cell simplifies to
\begin{equation}
\label{eqn:realspacelocalmodeR0}
\mathbf{\tilde{a}}^x_n(\mathbf{R_\lambda=\mathbf{0}}) =
\frac{1}{\sqrt{N_k}}
\sum_{\mathbf{k}}
\mathbf{a}^x_n(\mathbf{k}).
\end{equation}
This construction yields a displacement pattern that is spatially localized around the reference unit cell in the limit of $N_k \rightarrow \infty$. We note that where internal coordinate definitions span a unit cell boundary, a representation for the maximally localized $\mathbf{\tilde{a}}^x_n(\mathbf{R}_\lambda)$ would require a supercell construction.

\subsubsection{Properties of real-space-localized vibrational modes}
We recall that the adiabatic vector $\mathbf{a}^x_n(\mathbf{k})$ at wavevector $\mathbf{k}$ describes the linear elastic relaxation of the crystal along the normal modes in response to a periodic internal coordinate distortion. By analogy, a real-space localized adiabatic vector $\mathbf{\tilde{a}}^x_n(\mathbf{R}_\lambda)$ describes the overall relaxation of the infinite crystal that results from a localized distortion of $q_n$, obtained through the combined response of all available normal modes, at all wavevectors. The elastic response to an internal coordinate distortion associated with each wavevector $\mathbf{k}$ is quantified by the projected compliance,
\begin{equation}
k^a_n(\mathbf{k})^{-1}  = \mathbf{d}_n(\mathbf{k}) \; \mathbf{K}^{-1}(\mathbf{k}) \; \mathbf{d}_n^\dagger(\mathbf{k})
\end{equation}
which is both phase-invariant and additive when the system relaxes along multiple independent channels. Correspondingly, we propose that an effective force constant for $\mathbf{\tilde{a}}^x_n(\mathbf{R}_\lambda)$ can be obtained by
\begin{equation}
\label{eqn:realspaceforceconstant}
\tilde{k}^{a}_n =
\left[
\frac{1}{N_k}
\sum_{\mathbf{k}}
\mathbf{d}_n(\mathbf{k}) \;
\mathbf{K}^{-1}(\mathbf{k}) \;
\mathbf{d}_n^\dagger(\mathbf{k})
\right]^{-1},
\end{equation}
where the Brillouin-zone summation is performed over the projected inverse normal-mode force constants at each wavevector.

\Cref{eqn:realspaceforceconstant} takes the form of a simple harmonic mean of the wavevector-resolved force constants, which follows directly from the definition of the wavevector-resolved adiabatic vectors in \Cref{eqn:ax}  where the projected compliance $\mathbf{d}_n(\mathbf{k}) \; \mathbf{K}^{-1}(\mathbf{k}) \; \mathbf{d}_n^\dagger(\mathbf{k})$ appears as the normalization of the elastic response at each $\mathbf{k}$. Since the real-space localized adiabatic vector $\mathbf{\tilde{a}}^x_n(\mathbf{R}_\lambda)$ is constructed from the superposition of the independently relaxed wavevector-resolved responses, \Cref{eqn:realspaceforceconstant} emerges as the direct scalar analog of the Wannier-type construction in \Cref{eqn:realspacelocalmodeR0}.

Once $\tilde{k}^{a}_n$ are obtained, the corresponding local vibrational frequencies are given by
\begin{equation}
\tilde{\omega}_n^a = \sqrt{m_{n}^{-1} \; \tilde{k}^a_n}.
\end{equation}

\subsubsection{Symmetry-invariance of Wannier-type local vibrational mode properties}{\label{sec:symmetryBZ}}

The potential within a crystal is invariant under all point group symmetries of the lattice (symmetry group~$\mathcal{G}$). As a consequence, under a symmetry operation $g \in \mathcal{G}$, represented by the Cartesian transformation matrix $\mathcal{U}_g$, the matrix $\mathbf{f}^x(\mathbf{k})$ satisfies
\begin{equation} \label{eqn:fx_covariance}
    \mathbf{f}^x(g \mathbf{k}) = \mathcal{U}_g \; \mathbf{f}^x(\mathbf{k}) \; \mathcal{U}_g^{\mathsf T},
\end{equation}
with
\begin{equation} \label{eqn:Lk_covariance}
    \mathbf{L}(g \mathbf{k}) = \mathcal{U}_g \; \mathbf{L}(\mathbf{k})
\end{equation}
and
\begin{equation} \label{eqn:Kk_invariance}
    \mathbf{K}(g \mathbf{k}) = \mathbf{K}(\mathbf{k}).
\end{equation}

While the phonon eigenvalues in \Cref{eqn:Kk_invariance} are invariant under $g\in \mathcal{G}$, the same is not true of the normal mode eigenvectors in \Cref{eqn:Lk_covariance}. These eigenvectors transform covariantly and acquire a transformation in the atomic displacement space given by $\mathcal{U}_g$. We additionally find (see ESI~S2) that
\begin{equation}
\label{eqn:bcovar}
    \mathbf{b}_{gn}(\mathbf{k}) = \mathbf{b}_{n}(g^{-1} \mathbf{k}) \; \mathcal{U}_g^{\mathsf T},
\end{equation}
such that the construction of $\mathbf{D}(\mathbf{k})$ becomes invariant under the crystallographic point group symmetries only up to a permutation of internal-coordinate labels $n \mapsto gn$ within a symmetry orbit (\emph{i.e.,} it is covariant),
\begin{equation} \label{eqn:covariance_D}
\begin{aligned}
    \mathbf{d}_{gn}(\mathbf{k}) &= \mathbf{b}_{gn}(\mathbf{k}) \; \mathbf{L}(\mathbf{k})
    \\ &= \mathbf{b}_{n}(g^{-1} \mathbf{k}) \; \mathcal{U}_g^{\mathsf T} \; \mathcal{U}_g \; \mathbf{L}(g^{-1} \mathbf{k}) \\ 
    &= \mathbf{b}_{n}(g^{-1} \mathbf{k}) \; \mathbf{L}(g^{-1} \mathbf{k}) \\
    &= \mathbf{d}_{n}(g^{-1} \mathbf{k}).
\end{aligned} 
\end{equation}

Consequently, wavevector-resolved adiabatic vectors and projected compliances transform covariantly under point group symmetries. Specifically, for an internal coordinate $n$ mapped to internal coordinate $gn$ \textit{via} symmetry operation $g$, the projected compliance satisfies
\begin{equation} \label{eqn:covariance_compliance}
    k_{gn}^a(\mathbf{k})^{-1} = k_n^a(g^{-1} \mathbf{k})^{-1}.
\end{equation}
An analogous relation holds for the wavevector-resolved adiabatic vectors $\mathbf a^x(\mathbf{k})$, which transform as
\begin{equation} \label{eqn:covariance_axk}
    \mathbf{a}^x_{gn}(\mathbf{k}) = \mathcal{U}_g \; \mathbf{a}^x_{n}(g^{-1} \mathbf{k}),
\end{equation}
where $\mathcal{U}_g$ enters through the covariance of the $\mathbf{L}(\mathbf{k})$ matrix in \Cref{eqn:ax}. We note that for degenerate normal modes, symmetry may mix eigenvectors within degenerate subspaces. This gauge freedom does not, however, affect the symmetry-covariance of the projected quantities constructed from $\mathbf D(\mathbf k)$.

The Wannier-type local vibrational modes that are constructed from $\mathbf{a}^x(\mathbf{k})$ and $k^a(\mathbf{k})$ must obey the point group symmetry of the lattice. Correspondingly, $\mathbf{\tilde{a}}^x(\mathbf{R}_\lambda)$ and $\tilde{k}^a$ must transform as
\begin{equation} \label{eqn:covariance_aR}
\mathbf{\tilde{a}}^x_{g n}(\mathbf{R}_\lambda) = \mathcal U_g \; \mathbf{\tilde{a}}^x_n(\mathbf{R}_\lambda)
\end{equation}
and
\begin{equation} \label{eqn:invariance_ktilde}
\tilde{k}^a_{g n} = \tilde{k}^a_n.
\end{equation}
The covariance of wavevector-resolved local vibrational mode properties gives rise to important considerations for the practical implementation of our Wannier-type construction. In conventional lattice dynamics calculations, Brillouin‑zone integrations are performed over the irreducible Brillouin zone (IBZ), leveraging symmetry weighting to account for contributions from equivalent wavevectors. This reduction is formally valid for wavevector-dependent quantities that are invariant under the symmetry operations of the crystallographic point group, such as phonon eigenvalues or other scalar properties satisfying $f(\mathbf{k}) = f(g\mathbf{k})$ for all $g \in \mathcal{G}$. As we have demonstrated in \Cref{eqn:covariance_D,eqn:covariance_compliance,eqn:covariance_axk}, this relation does not apply to the projected wavevector-dependent quantities that form the basis of our Wannier-type construction. 

\begin{figure}[hbtp!]
    \centering
    \includegraphics[width=.9\linewidth]{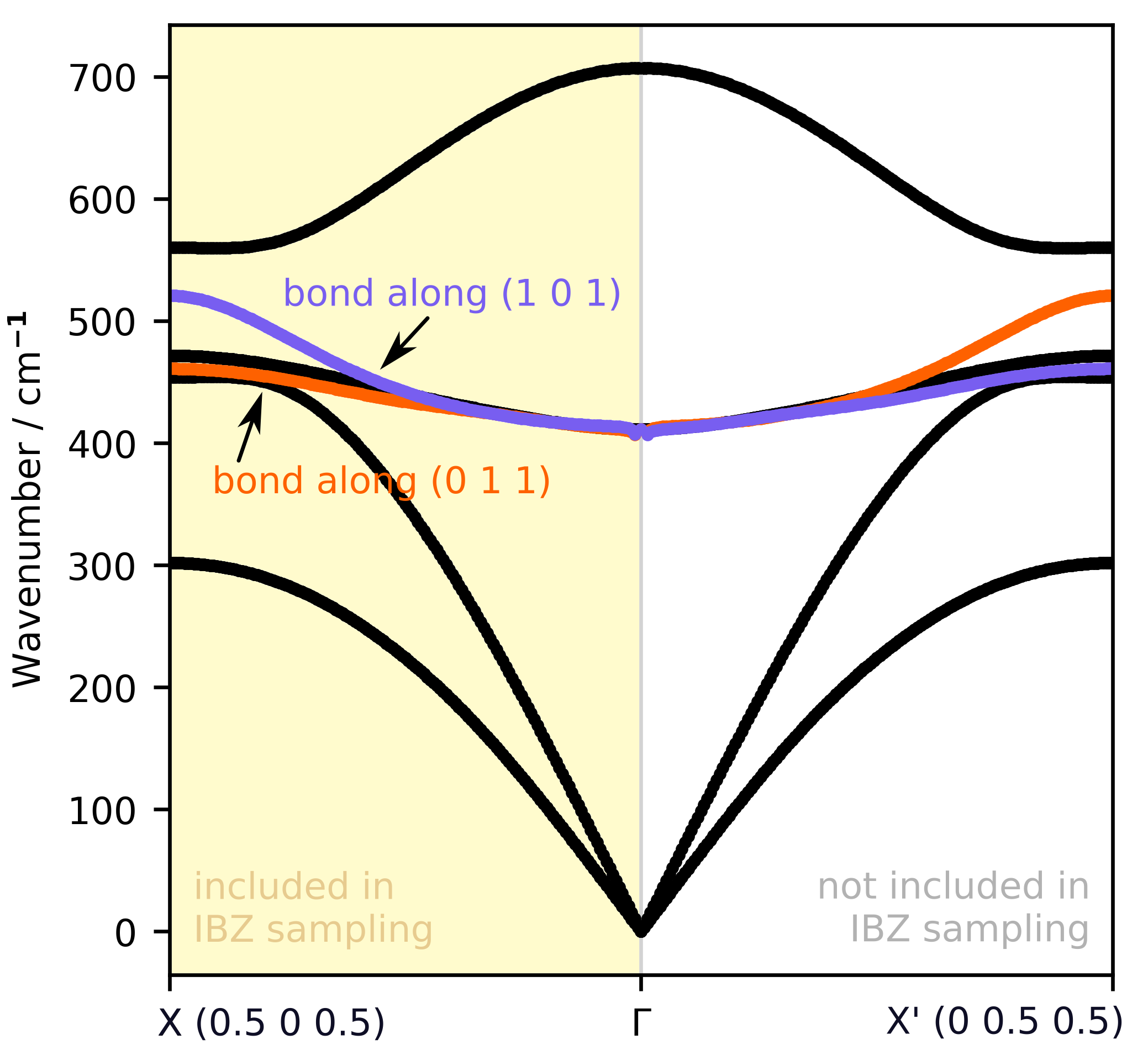}
    \caption{Covariance of wavevector-resolved local-mode frequencies in \ce{MgO}. The normal-mode dispersion (black) is given alongside the local-mode dispersion (purple, orange).  The \ce{Mg-O} bond $n$ oriented along $[1,0,1]$ exhibits different local-mode frequencies (purple) at symmetry-related k-points $X=(0.5,0,0.5)$ and $X'=g^{-1}X=(0,0.5,0.5)$, related by $g=C_2$ symmetry about the $[110]$ axis. The symmetry-related bond $gn$, oriented along $[0,1,1]$, exhibits the opposite behavior. This is a direct consequence of the covariance relation $k_{gn}^a(\mathbf{k})^{-1} = k_n^a(g^{-1} \mathbf{k})^{-1}$.}
    \label{fig:covariance}
\end{figure}

By sampling the IBZ representation, we can capture all unique local-mode behaviors. However, individual members of an internal-coordinate symmetry orbit, in general, behave differently in the irreducible wedge, \Cref{fig:covariance}. As a consequence, when the Brillouin zone is sampled only over its irreducible representation, contributions from symmetry‑related wavevectors are not correctly included in the Brillouin-zone sums in \Cref{eqn:realspacelocalmode,eqn:realspaceforceconstant}. The symmetry covariance relations in \Cref{eqn:covariance_compliance,eqn:covariance_axk} are therefore not explicitly represented within an IBZ-only sampling, such that the required relations in \Cref{eqn:covariance_aR,eqn:invariance_ktilde} are not enforced.

To perform a Wannier-type construction of $\mathbf{\tilde{a}}^x(\mathbf{R}_\lambda)$ and the associated $\tilde{k}^a$, we must therefore consider a symmetry‑complete (reducible) sampling of the Brillouin zone by explicitly evaluating wavevector-resolved local modes for all symmetry‑related wavevectors. This can be achieved by sampling the Brillouin zone (RBZ) on a complete $\Gamma$-centered Monkhorst-Pack grid, or by reconstructing the star of each wavevector from the IBZ prior to the localization procedure. Only in this way is the full point‑group symmetry of the lattice response correctly recovered in $\mathbf{\tilde{a}}^x(\mathbf{R}_\lambda)$ and $\tilde{k}^a$.

In this work, we chose to compute dynamical matrices on complete Monkhorst-Pack grids during the lattice dynamics calculation. This choice was made to ensure consistent numerical precision in the input data for the construction of wavevector-resolved local vibrational modes. If one instead chooses to reconstruct the full Brillouin zone sampling from the IBZ, the rows of the $\mathbf{D}(g^{-1}\mathbf{k})$ matrices, \Cref{eqn:dmat}, are obtained as
\begin{equation}
\begin{aligned}
    \mathbf{d}_n(g^{-1} \mathbf{k}) &= \mathbf{b}_n(g^{-1} \mathbf{k}) \; \mathbf{L}(g^{-1} \mathbf{k}) \\
    &= \mathbf{b}_n(g^{-1} \mathbf{k}) \; \mathcal{U}_g^{\mathsf T} \; \mathbf{L}(\mathbf{k}),
\end{aligned}
\end{equation}
using the covariance of the normal-mode eigenvectors in \Cref{eqn:Lk_covariance}. 

\subsection{Comment on the physical interpretation of wavevector-resolved and Wannier-type local vibrational modes}{\label{sec:interpretation}}

Our Wannier-type local vibrational modes describe internal-coordinate vibrations that are spatially localized around a single reference unit cell. This contrasts with the wavevector-resolved local modes, which are localized to a specific internal coordinate in every unit cell, with the relative phase of their distortion being defined by the wavevector. It follows from this difference in physical picture that the respective adiabatic vectors also have different physical interpretations. The Wannier-type $\mathbf{\tilde{a}}^x(\mathbf{R}_\lambda)$ describe the full elastic response of the crystal to an individual internal coordinate distortion, allowing the crystal to relax through all the available normal mode channels at all available wavevectors. Instead, the wavevector-resolved $\mathbf{a}^x(\mathbf{k})$ describe how the crystal relaxes only \textit{via} the normal mode channels at the wavevector for which it is calculated.

In the context of the present manuscript, we must also consider the difference in the local mode force constants that are obtained through these two approaches. Using the Wannier-type construction, the local mode force constants $\tilde{k}^a$ represent the resistance of the entire crystal to a localized internal-coordinate distortion. As these force constants are constructed from the compliance matrices at all sampled wavevectors, they explicitly account for the phonon dispersion behavior, \Cref{eqn:realspaceforceconstant}. Hence, $\tilde{k}^a$ account for a complete representation of the crystal response to a particular local deformation. In comparison, the wavevector-resolved force constants $k^a(\mathbf{k})$ describe the stiffness in response to a Bloch-periodic internal-coordinate distortion relaxed only \emph{via} the normal modes with the same periodicity. The terminology and notation used to describe the local modes in this work as well as their physical interpretation are summarized in \Cref{tab:notation}.

\begin{table*}[htbp]
\centering
\caption{Summary of the types of local modes discussed in this work. The adiabatic vectors represent the relaxation of the entire crystal in response to the described internal coordinate distortions, and the corresponding force constants quantify the associated stiffness.}
\label{tab:notation}
\begin{tabular}{c|ccc|c}
Type of local mode & Description & Adiabatic vector & Force constant &  Interpretation \\ \hline
\multirow{2}{*}{Wavevector-resolved} & $\Gamma$-point & $\mathbf{a}^x(\mathbf{k}=\Gamma)$ &  $k^a(\mathbf{k}=\Gamma)$ & \makecell[l]{Periodic internal coordinate distortion repeated \\ identically in every unit cell.} \\
                                     & Finite-wavevector & $\mathbf{a}^x(\mathbf{k})$ & $k^a(\mathbf{k})$ & \makecell[l]{Periodic internal coordinate distortion with \\ periodicity determined by the wavevector $\mathbf{k}$.} \\ \hline
Wannier-type                         & Real-space localized & $\tilde{\mathbf{a}}^x(\mathbf{R}_\lambda)$ & $\tilde{k}^a$  & \makecell[l]{Individual internal coordinate distortion localized \\ on a single unit cell.}
\end{tabular}
\end{table*}

The delocalised nature of the wavevector-resolved local modes makes them a natural tool for analyzing the contributions of structural and chemical features to the phonon dispersion behavior,\cite{mojsak_jctc_2026} or for investigating cooperative distortions across periodic lattices. Wannier-type local modes are instead more closely aligned with the physical picture of deforming an individual interaction within the crystal. Consequently, the Wannier-type modes $\tilde{\mathbf{a}}^x(\mathbf{R}_\lambda)$ and force constants $\tilde{k}^a$ emerge as the natural objects to compare directly with local vibrational modes in isolated molecules, and to interpret as bond or interaction strength descriptors for solids.

\section{Computational Methods}

All geometry optimization and phonon calculations were performed within the framework of plane-wave density functional theory (DFT) as implemented in CASTEP v25.11.\cite{clark_first_2005} Input crystal structures were obtained from the Inorganic Crystal Structure Database (ICSD) with collection codes 159375 (MgO),\cite{hazen1976effects} 656475 (diamond),\cite{asanoNeutronPowderDiffraction1988} 52457 (silicon),\cite{straumanisLatticeParametersCoefficients1952} 603798 (silicon carbide),\cite{liThermalExpansionCubic1986} 52152 (aragonite) and 52151 (calcite).\cite{pilatiLatticeDynamicalEstimationAtomic1998} The structure of the isolated carbonate anion was purpose-built with equal bond lengths and angles, and placed in a cubic $18\times18\times18$~\AA{} unit cell, to minimize interactions between periodic images, see ESI~S3.7 for further detail. The orientation of the ion in the unit cell was chosen to preserve its three-fold rotational axis and in-plane two-fold axes relative to the unit cell symmetry (the coordinates are given in ESI~S3.7). This was done to maximize the molecular point group symmetries that are explicitly preserved and enforced in our representation. A uniform compensating background charge of $-2$ was included to neutralize the unit cell and allow the Coulomb potential to converge under periodic boundary conditions.

The input structures were relaxed with the exchange--correlation energy described using the generalized gradient approximation (GGA) functional of Perdew-Burke-Ernzerhof (PBE)\cite{perdew_generalized_1996}. The Grimme D2 dispersion correction\cite{grimme_semiempirical_2006} was used for all systems except for the carbonate anion. The nuclear Coulomb potential was attenuated with norm-conserving pseudopotentials as obtained on-the-fly within CASTEP. The wavefunction was expanded in a plane-wave basis set to a kinetic energy cut-off of 1200~eV for \ce{MgO} and the \ce{CaCO3} polymorphs and 900~eV for \ce{C}, \ce{SiC} and \ce{Si}, and sampled on a Monkhorst-Pack k-point grid with spacing of 0.05~\AA$^{-1}$. Convergence criteria for the self-consistent field cycles were set to an electronic energy change $<10^{-10}$~eV and an electronic eigenvalue change $<10^{-12}$~eV. The Broyden density-mixing scheme was used to accelerate SCF convergence, with a mixing amplitude of 0.5 and a maximum mixing G-vector of 1.5~\AA$^{-1}$. Structure relaxation used the limited-memory Broyden-Fletcher-Goldfarb-Shanno (LBFGS) algorithm,\cite{LBFGS} with convergence reached when the total energy change was $<2\times10^{-6}$~eV~atom$^{-1}$, residual forces $<10^{-4}$~eV~\AA$^{-1}$, ionic displacements $<10^{-5}$~\AA{} and cell strain $<0.1$~GPa. Further detail regarding the input and optimized structures is provided in the ESI,~S3.

Phonon calculations were performed using the linear response method, as implemented in CASTEP v25.11,\cite{refson_prb_2006} with LO-TO splitting enabled for MgO, SiC and \ce{CaCO3}, and disabled for the remaining systems: diamond, silicon and the carbonate ion. Brillouin zones of the model systems were sampled on $\Gamma$-centered Monkhorst-Pack grids with \emph{ca}~0.05~\AA$^{-1}$ k-point spacing, and only at the $\Gamma$-point for the carbonate ion. For the production of the phonon dispersion relations, dynamical matrices were interpolated from these grids onto paths between high-symmetry Brillouin zone points, as generated by the SeeK-path utility.\cite{seekpath, togo_spglib_2024} For the Wannier-type localization procedure, dynamical matrices were computed on Monkhorst-Pack grids as specified in the main text. To obtain explicit, complete representations of the required Brillouin zone grids in the CASTEP output rather than their irreducible subsets, the k-point coordinates in the Monkhorst-Pack grids were provided using the \verb|PHONON_KPOINT_LIST| or the \verb|PHONON_FINE_KPOINT_LIST| blocks in the \verb|.cell| input file. For illustrative purposes, we highlight in ESI~S4 the computational costs of performing the phonon dispersion calculations reported in this work.

Wavevector-resolved local modes were computed for the internal coordinates described in the main text, according to the formalism described in the Theoretical Background using our LModeA-k software.\cite{mojsak_jctc_2026} The Wannier-type local mode construction and the calculation of the associated properties was done using a post-processing script, available at our group GitHub (see Data Availability Statement accompanying this manuscript).

\section{Results and Discussion}

Here we demonstrate the construction and application of Wannier-type local modes for a series of model systems. With the first example of \ce{MgO}, we show how an increasingly dense sampling of the Brillouin zone leads to converged Wannier-type local modes $\mathbf{\tilde{a}}^x(\mathbf{R}_\lambda)$ with real-space localized vibrational behavior that are accompanied by converged force constants $\tilde{k}^a$. We apply our procedure to a set of isostructural tetrahedrally-coordinated covalent networks, demonstrating that Wannier-type local mode force constants are reliable metrics for intrinsic bond strengths in crystals. Finally, by studying two polymorphs of \ce{CaCO3}, we explore how our Wannier-type local modes capture the influence of crystal packing on the strengths of intramolecular covalent bonds in the \ce{CO3^2-} anion and on the intermolecular \ce{CO3^2-}$\cdots$\ce{Ca^2+} interactions. 

\begin{figure}[hbtp!]
    \centering
    \includegraphics[width=.9\linewidth]{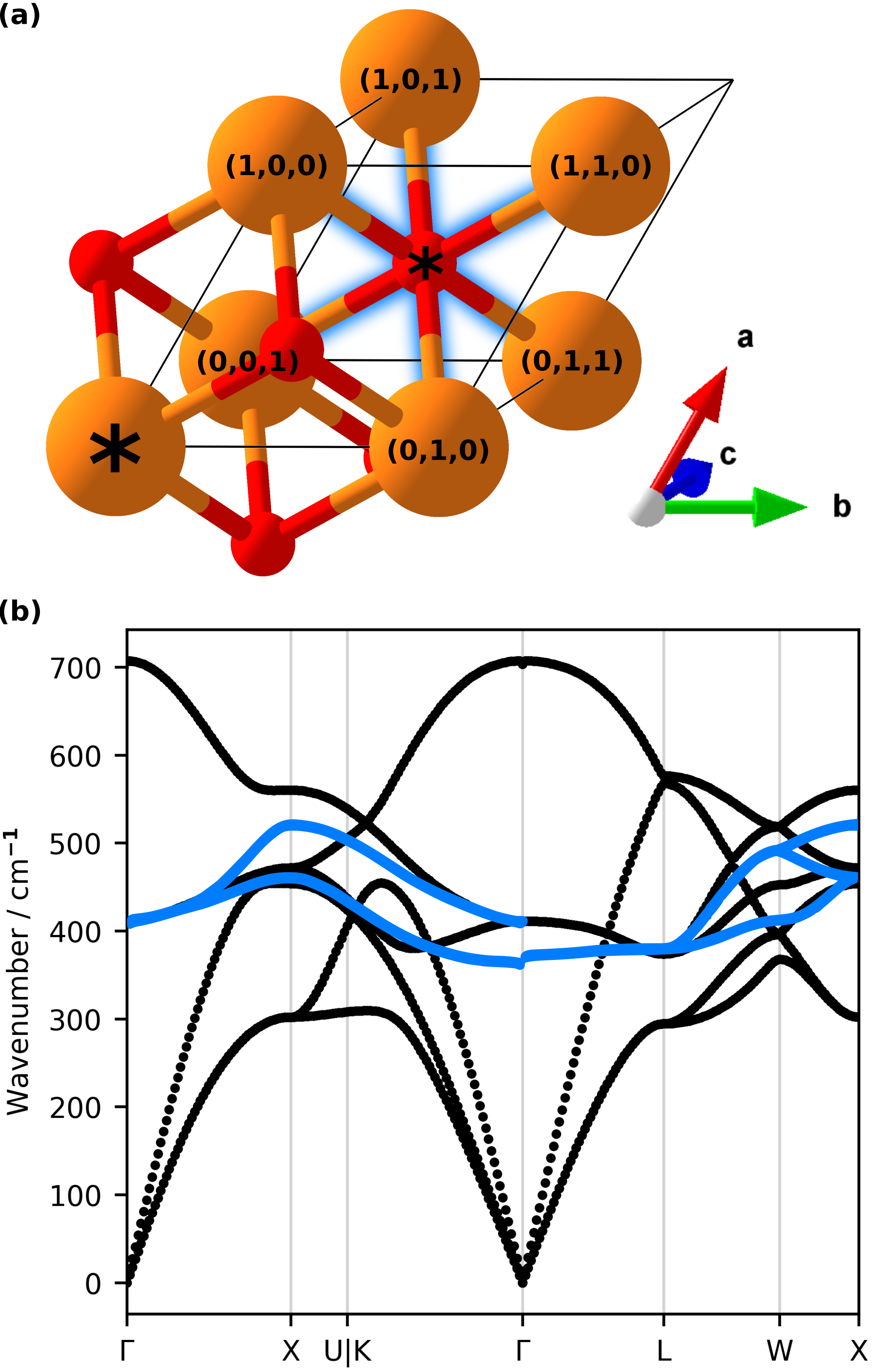}
    \begin{subfigure}{0pt}
        \phantomcaption
        \label{fig:mgo:struct}
    \end{subfigure}
    \begin{subfigure}{0pt}
        \phantomcaption
        \label{fig:mgo:disp}
    \end{subfigure}
    \caption{Structure and phonon dispersion relations of \ce{MgO}. (a) The primitive unit cell structure of \ce{MgO} (black lines). Ions are colored as: \ce{Mg} (orange), \ce{O} (red). Ions indicated with asterisks formally belong to the displayed primitive cell. Bonds used in our analysis are highlighted in blue and the unit cells to which the \ce{Mg} ions belong are indicated atop their representations. (b) Phonon dispersion curves of the normal mode frequencies (black) and the dispersion behavior of the six \ce{Mg-O} bond stretch local modes (blue) as a function of wavevector.}
    \label{fig:mgo}
\end{figure}

\begin{figure*}[htbp!]
    \centering
    \includegraphics[width=.8\linewidth]{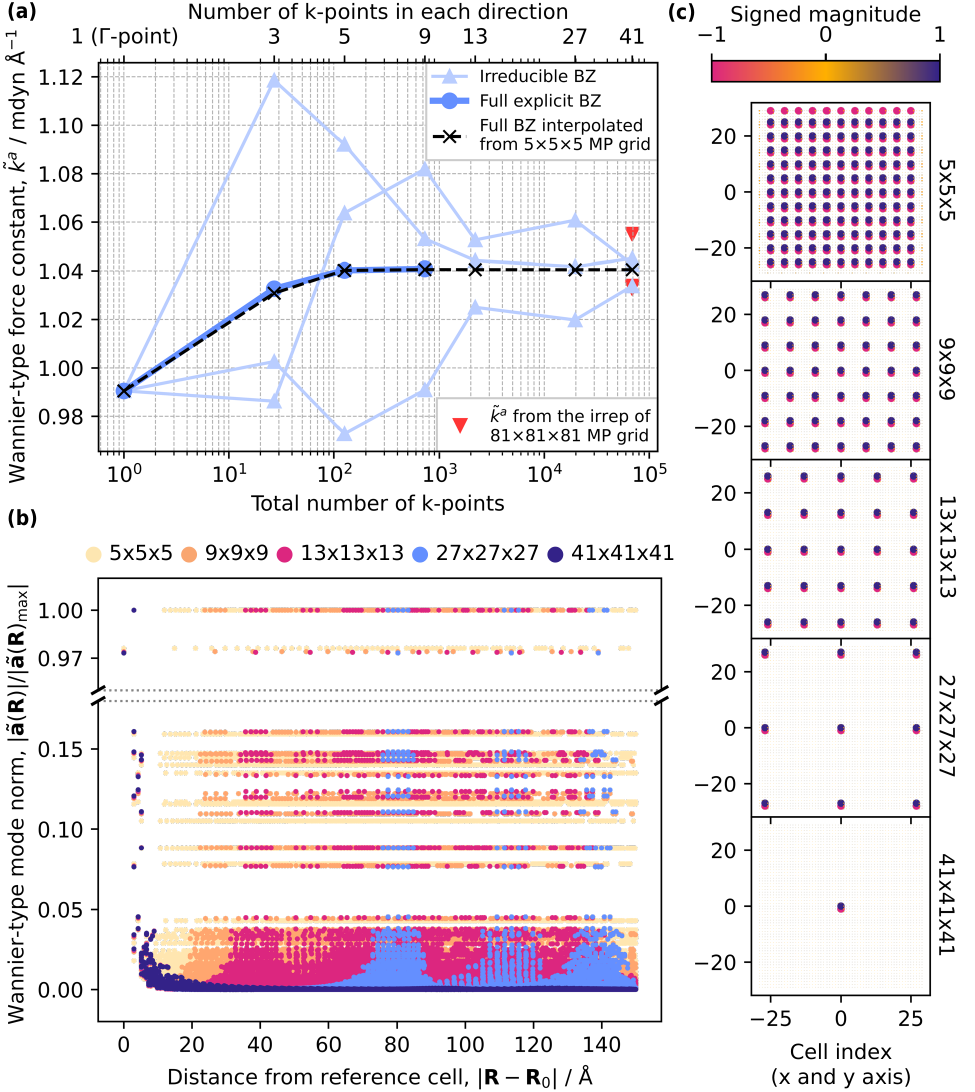}
    \begin{subfigure}{0pt}
        \phantomcaption
        \label{fig:conv:fcs}
    \end{subfigure}
    \begin{subfigure}{0pt}
        \phantomcaption
        \label{fig:conv:norms}
    \end{subfigure}
    \begin{subfigure}{0pt}
        \phantomcaption
        \label{fig:conv:mags}
    \end{subfigure}
    \caption{Convergence behavior of $\tilde{k}^{a}$ and $\mathbf{\tilde{a}}^x(\mathbf{R}_\lambda)$ in \ce{MgO}. (a) Convergence of $\tilde{k}^{a}$ for symmetry-equivalent \ce{Mg-O} local modes as the k-point sampling density is increased. The convergence of $\tilde{k}^a$ computed from sampling the RBZ on Monkhorst-Pack (MP) grids up to $9\times9\times9$ is shown as a solid blue line. Larger MP grids were obtained by interpolating from a $5\times5\times5$ grid. Values of $\tilde{k}^a$ obtained from these interpolated grids are given as a dotted black line. The behavior of $\tilde{k}^a$ when generated from sampling the IBZ only is shown as light blue lines, with the values computed from a $81\times81\times81$ MP grid shown in red, highlighting that increased sampling of the IBZ does not reproduce the correct behavior. (b) Real-space decay of Wannier-type local mode norms for the Mg(0,0,1)--O(0,0,0) stretch in unit cells outwith $\mathbf{R}_\lambda = \mathbf{0}$, calculated by sampling the RBZ using different k-point MP grids. (c) Signed magnitude of the real-space mode norms for the Mg(0,0,1)--O(0,0,0) stretch in the $bc$ plane. Magnitude is represented by the dot size, and sign relative to the Wannier-type mode at $\mathbf{R}_\lambda = \mathbf{0}$ by the dot color. Neighboring oppositely-colored dots correspond to $\mathbf{\tilde{a}}^x(\mathbf{R}_\lambda)$ norms distributed across neighboring unit cells due to the formal assignment of ions defining the internal coordinates to different unit cells.}
    \label{fig:conv}
\end{figure*}

\subsection{Convergence of Wannier-type local vibrational modes with Brillouin zone sampling density}

We first consider the ion-pair interactions in the simple rock-salt crystal of \ce{MgO} (space group \textit{Fm}$\bar{3}$\textit{m}), \Cref{fig:mgo:struct}. The primitive cell of \ce{MgO} contains six symmetry-equivalent \ce{Mg-O} interactions. Though equivalent by symmetry, for our purposes it is illustrative to consider them individually. These \ce{Mg-O} interactions are defined by bonds connecting the \ce{O} ion at the center of the reference cell, $\mathbf{R_\lambda} = \mathbf{0}$, to the \ce{Mg} ions that sit on the corners of the primitive cell. As such, these \ce{Mg} ions formally reside in neighboring cells: $(0,0,1)$, $(0,1,0)$, $(0,1,1)$, $(1,0,0)$, $(1,0,1)$, and $(1,1,0)$. While the six selected internal coordinate representations involve ions that formally belong to different primitive cells, the internal coordinates themselves (\textit{i.e.,} the bonds) are fully confined in the reference cell.

Akin to the normal-mode dispersion behavior, the six \ce{Mg-O} bond-stretch local modes exhibit a significant degree of dispersion across the Brillouin zone, \Cref{fig:mgo:disp}. 
At the $\Gamma$-point, all six local modes are degenerate when LO-TO splitting is neglected, consistent with the full cubic symmetry of the crystal. This degeneracy is lifted away from $\Gamma$, with the local mode behaviors splitting into symmetry-related subsets whose composition depends on the orientation of $\mathbf{k}$. Along the $\Delta (\Gamma\!-\!X)$, $\Sigma (\Gamma\!-\!K)$ and $X\!-\!U$ paths (which all traverse two reciprocal space lattice vectors symmetrically) the real-space projection of four \ce{Mg-O} bond vectors onto the wavevector direction are equivalent. Correspondingly, these four \ce{Mg-O} local modes remain degenerate along these high-symmetry paths, whilst the remaining two bonds form a second symmetry-degenerate pair. In contrast, along the $W\!-\!X$ path, which samples the Brillouin zone asymmetrically with respect to all three reciprocal lattice vectors, the six \ce{Mg-O} bonds split into three pairs of symmetry-degenerate local modes.

Given the significant dispersion seen in \Cref{fig:mgo:disp}, we find that the local mode frequency at $\mathbf{k}=\Gamma$ is a poor reflection for the average mode frequency for all six \ce{Mg-O} local modes across the Brillouin zone. In fact, the local mode frequencies calculated at any finite wavevector for \ce{Mg-O} bonds are typically higher than at $\Gamma$. Correspondingly, the local mode force constant calculated at $\Gamma$, $k^a(\mathbf{k}=\Gamma) = 0.991$~mdyn~\AA$^{-1}$ (in the absence of LO-TO splitting) is also a poor reflection of the average local mode force constant for the \ce{Mg-O} interactions.

To explore how the local modes are affected by the incorporation of phonon dispersion into their construction, we subsequently computed $\mathbf{\tilde{a}}^x(\mathbf{R_\lambda})$, \Cref{eqn:realspacelocalmode}, and the corresponding $\tilde{k}^{a}$, \Cref{eqn:realspaceforceconstant}, for each of the six \ce{Mg-O} bonds by sampling the reducible Brillouin zone (RBZ; see discussion in Section~\ref{sec:symmetryBZ}). As we systematically increased the sampling density of the RBZ, \Cref{fig:conv:fcs}, the resulting $\tilde{k}^{a}$ converged to 1.040~mdyn~\AA$^{-1}$, varying by no more than 0.001~mdyn~\AA$^{-1}$ once a Monkhorst-Pack k-point grid of $5\times5\times5$ grid was reached (corresponding to \emph{ca} 0.10~\AA$^{-1}$ k-point spacing). This converged value contrasts that obtained by sampling only the $\Gamma$-point, $k^a(\mathbf{k}=\Gamma) = 0.991$~mdyn~\AA$^{-1}$ (\emph{ca}~$5\%$ smaller). The origins of this discrepancy are two-fold: by using a Wannier-type local mode construction, (1) the adiabatic vectors are no longer periodic, and (2) they reflect how the entire periodic system, through normal modes at all wavevectors, relaxes when the internal coordinate is perturbed. In the case of \ce{MgO} this description reveals a markedly stiffer system response than that described by the local modes defined at the $\Gamma$-point.

We additionally calculated $\tilde{k}^{a}$ from a symmetry-weighted sampling of the irreducible Brillouin zone (IBZ). This sampling approach led to large fluctuations in the predicted values of $\tilde{k}^{a}$, which did not converge with improved IBZ sampling density, \Cref{fig:conv:fcs}. Moreover, even a large sampling density of the IBZ yielded three independent values for $\tilde{k}^{a}$, each representing a subset of the six \ce{Mg-O} local modes. Notably, all three of these values differ from the true converged $\tilde{k}^{a}$. This result demonstrates decisively the need to explicitly sample the RBZ when constructing Wannier-type local vibrational modes (see Section~\ref{sec:symmetryBZ} for further discussion).

To illustrate that our converged $\tilde{k}^a$ correspond to spatially-localized vibrations, we compute $\mathbf{\tilde{a}}^x(\mathbf{R}_\lambda)$, \Cref{eqn:realspacelocalmode}, across a $60\times60\times60$ supercell and obtain the norms of $\mathbf{\tilde{a}}^x(\mathbf{R}_\lambda)$ in each constituent unit cell. These norms are the largest in the reference cell at $\mathbf{R}_\lambda = \mathbf{0}$ and its nearest neighbors, since the adiabatic vectors are localized on internal coordinates defined by ions that formally belong to different unit cells. As the distance from the reference cell increases, the mode norms decay, \Cref{fig:conv:norms}. This decay is faster and leads to smaller norms away from the reference cell as the RBZ sampling density is increased.

Since our method relies on a discrete sampling of wavevectors, \Cref{eqn:realspacelocalmode} represents a discrete Fourier transform over an $N_a \times N_b \times N_c$ Monkhorst-Pack grid. As in any finite‑grid Fourier representation, the resulting real‑space adiabatic vector is strictly periodic over a supercell of dimensions $N_a a \times N_b b \times N_c c$, where $a,b,c$ are the primitive-cell lattice constants. Consequently, the maximally localized displacement pattern centered around $\mathbf{R}_\lambda=\mathbf{0}$ is accompanied by periodic images (aliasing replicas) separated by these supercell vectors, as seen in \Cref{fig:conv:norms,fig:conv:mags}. These Fourier replicas are an inherent consequence of the finite k-point sampling and vanish in the limit $N_k\rightarrow\infty$. However, as seen by the convergence of $\tilde{k}^a$, the coupling of these Fourier replicas through the lattice diminishes as the RBZ sampling is improved. This yields well-converged force constants using grids as coarse as $5\times5\times5$. Thus, in direct analogy to similar Wannier-type localization procedures, our approach exhibits systematic and reliable convergence toward a truly localized internal-coordinate distortion as the density of sampled wavevectors increases.

\subsection{Wannier-type local mode force constants as bond strength descriptors}

\begin{figure*}[htbp!]
    \centering
    \includegraphics[width=\linewidth]{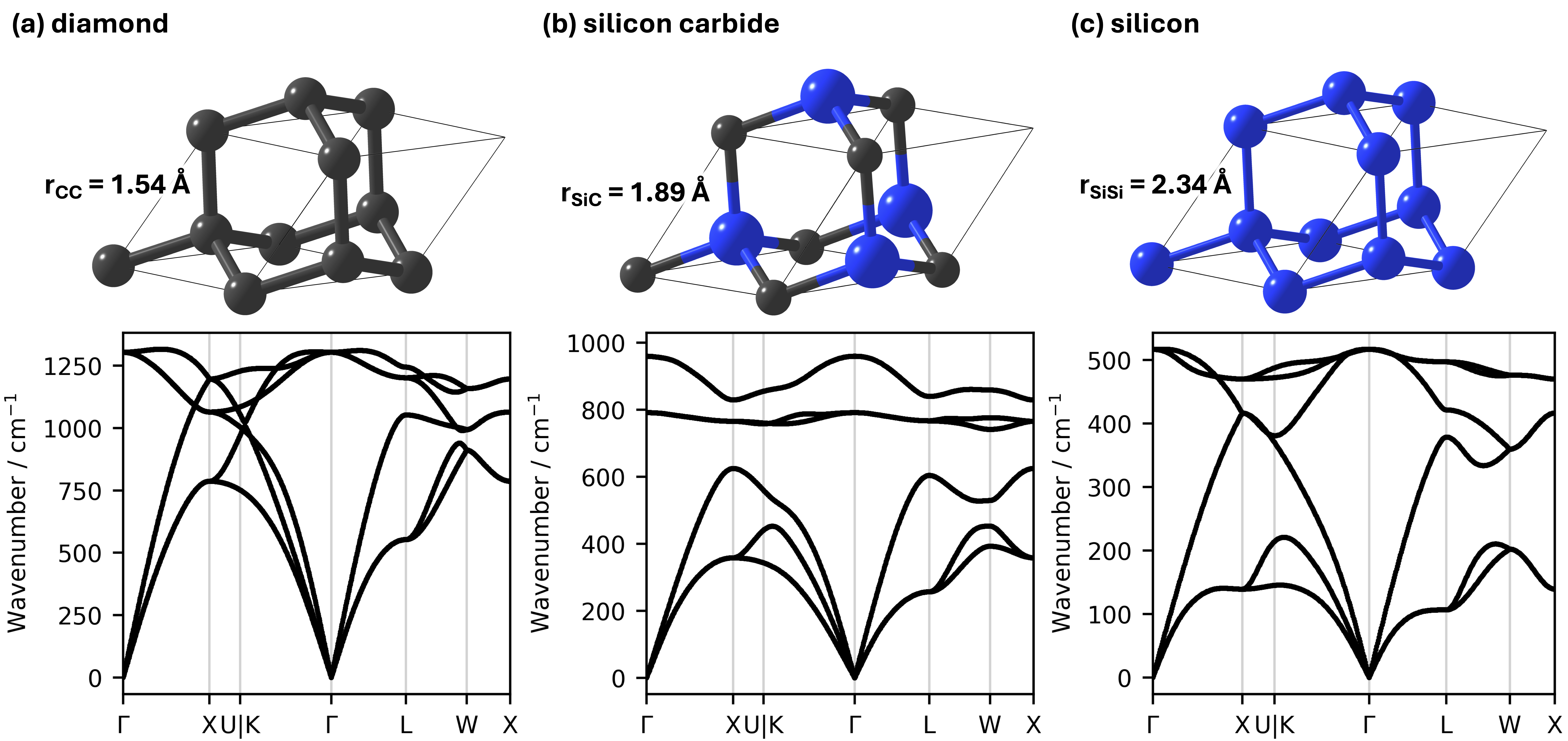}
    \begin{subfigure}{0pt}
        \phantomcaption
        \label{fig:diam:C}
    \end{subfigure}
    \begin{subfigure}{0pt}
        \phantomcaption
        \label{fig:diam:SiC}
    \end{subfigure}
    \begin{subfigure}{0pt}
        \phantomcaption
        \label{fig:diam:Si}
    \end{subfigure}
    \setlength{\abovecaptionskip}{-6pt}
    \caption{Crystal structures (top) and phonon dispersion relations (bottom) for: (a) diamond, (b) zinc blende-type silicon carbide and (c) diamond-type silicon. Primitive unit cell boundaries are drawn as black wires, with carbon atoms shown as black and silicon atoms shown as blue. Unit cells are not drawn to scale, with the nearest-neighbor bond lengths indicated alongside the structure diagrams for comparison. Note that the y-axis scales in the phonon dispersion plots are different.}
    \label{fig:diam}
\end{figure*}

We next explore the use of Wannier-type local modes to describe the covalent bonds in a family of tetrahedrally-coordinated covalent network solids: diamond, \ce{SiC}, and \ce{Si} (space groups \textit{Fd}$\bar{3}$\textit{m}, \textit{F}$\bar{4}3${m}, and \textit{Fd}$\bar{3}$\textit{m}, respectively), \Cref{fig:diam}. As reference, we compare the trends in our computed values of $\tilde{k}^a$ against experimentally determined cohesive energies for each crystal, with bond strength decreasing as \ce{C-C} $>$ \ce{Si-C} $>$ \ce{Si-Si}, \Cref{tab:bond_comparison}.

\begin{table}[htbp!]
\centering
\caption{Comparison of experimental and computed bond lengths and bond energies / Wannier-type local mode force constants for diamond (\ce{C}), \ce{SiC} and \ce{Si}. Experimental bond energies were estimated as half the cohesive energy per atom of the corresponding solid, taken from Ref.~\citenum{cohesive_energies}. The MP grids used to sample the RBZ for calculating $\tilde{k}^a$ were $33\times33\times33$ for diamond, $27\times27\times27$ for \ce{SiC}, and $21\times21\times21$ for \ce{Si}, corresponding in each case to \emph{ca} 0.015 \AA$^{-1}$ k-point spacing.}

\label{tab:bond_comparison}
\begin{tabular}{l|cc|ccc}
 & 
\multicolumn{2}{c}{Experimental} & 
\multicolumn{3}{c}{Computed} \\
\cline{2-3} \cline{4-6}
& 
\makecell{Bond \\ \hspace{1pt} length \hspace{1pt} \\ / \AA} &
\makecell{Bond \\ \hspace{1pt} energy \hspace{1pt} \\ / eV} &
\makecell{Bond \\ \hspace{1pt} length \hspace{1pt} \\ / \AA} &
\makecell{Wannier-type \\ $\tilde{k}^a$ \\ / mdyn\,\AA$^{-1}$} &
\makecell{$\Gamma$-point \\ $k^a(\mathbf{k}=\Gamma)$ \\ / mdyn\,\AA$^{-1}$}\\
\hline
\ce{C}         & 1.54 & 3.69 & 1.54 & 4.22 & 6.01  \\
\ce{SiC} & 1.89 & 3.18 & 1.89 & 2.61 & 3.48 \\
\ce{Si}         & 2.35 & 2.32 & 2.34 & 1.58 & 2.21 \\
\end{tabular}
\end{table}

Whether one considers $\tilde{k}^a$ by sampling across the RBZ or $k^a(\mathbf{k}=\Gamma)$ by sampling the $\Gamma$-point only, the same qualitative trend in bond strengths is observed, consistent with the trend in experimental cohesive energies. However, compared to $\tilde{k}^a$, the $\Gamma$-only results systematically overestimate the bond stiffness by 33--42\%. We can rationalize this overestimation by the fact that the optical phonon branches in these three crystals reach their maxima at $\mathbf{k}=\Gamma$ and soften at finite wavevectors. As a result, local mode force constants evaluated solely at the Brillouin-zone center probe the stiffest possible elastic response associated with periodic distortions of the covalent bond, rather than capturing the full elastic response of the system to individual bond distortions. This effect is the most pronounced in diamond (C), which shows the greatest degree of phonon dispersion.

By correctly describing the isolated covalent bond distortion through our Wannier-type procedure, we obtain values for the force constants that are significantly lower, and hence indicate a more deformable covalent bond network than might be expected from a delocalised $\Gamma$-point perspective. These results highlight an important consideration when using $\Gamma$-point-only local-mode analyses for bond characterization in extended solids. While $\Gamma$-point force constants can offer an indication of relative bond strengths, they should be used with caution when comparing materials that have different phonon dispersion behavior. This limitation is a direct consequence of the real-space interpretation of these local modes (see Section~\ref{sec:interpretation}).

\subsection{Sensitivity of Wannier-type local mode force constants to local chemical environments}

\begin{figure}[]
    \centering
    \includegraphics[width=\linewidth]{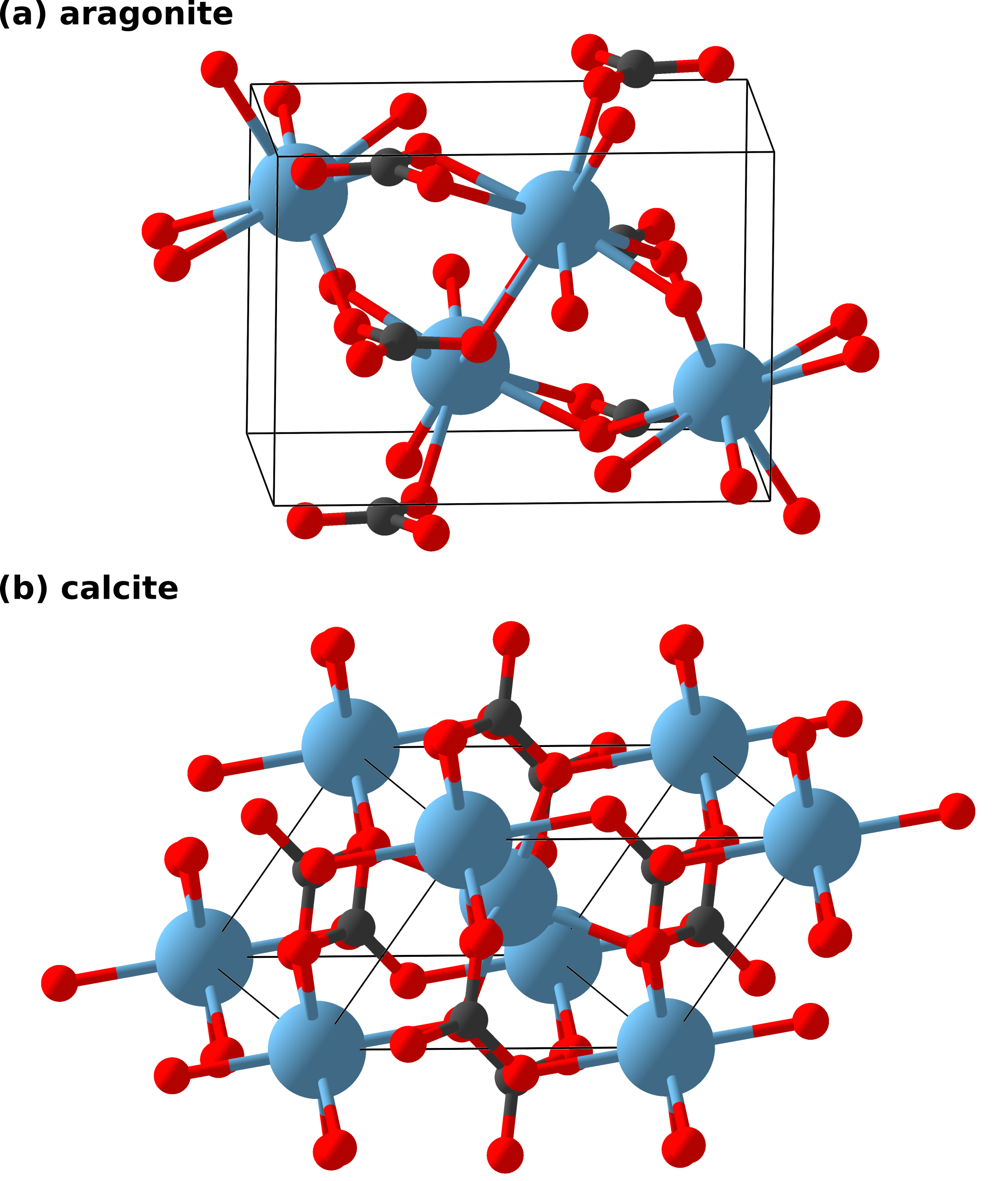}
    \caption{Crystal structures of two polymorphs of \ce{CaCO3}: (a) aragonite and (b) calcite. Primitive unit cell boundaries are shown as black wires with ions colored as: \ce{Ca^2+} (blue), \ce{C} (gray) and  \ce{O^2-} (red).}
    \label{fig:caco3}
\end{figure}

As a final illustration we explore the use of $\tilde{k}^a$ to describe subtle differences in local chemical environments that arise in polymorphic systems. To this end we consider the \ce{CaCO3} polymorphs aragonite (space group \textit{Pnma}) and calcite (space group \textit{R}$\bar{3}$\textit{c}), \Cref{fig:caco3}. While both structures contain quasi-planar carbonate anions, the coordination sphere of the calcium cation differs markedly: \ce{Ca^2+} is six-coordinate in calcite and nine-coordinate in aragonite. These differences provide a platform to assess whether Wannier-type force constants can resolve environment-dependent variations in the covalent \ce{C-O} bonding and in the predominantly ionic \ce{Ca^2+}$\cdots$\ce{O^2-} interactions. We additionally compare the force constants for the \ce{C-O} bonds in both polymorphs to those in an isolated \ce{CO3^2-} anion as a means to identify whether $\tilde{k}^a$ can capture the influence of crystal field effects on molecular bonding.

For both of the \ce{CaCO3} structures, convergence of the $\tilde{k}^a$ values required approximately 100 k-points to be sampled in the RBZ, corresponding to a $5\times5\times3$ Monkhorst-Pack grid for aragonite and a $5\times5\times5$ grid for calcite. While this grid number is similar to that needed to achieve convergence of $\tilde{k}^a$ in \ce{MgO}, it corresponds to a k-point spacing of \emph{ca} $0.05$~\AA$^{-1}$ compared to $0.10$~\AA$^{-1}$ in \ce{MgO}. This suggests that, while convergence can be achieved with relatively low sampling density for inorganic solids that exhibit notable phonon dispersion behavior, it should be assessed on a case-by-case basis for different systems.

\begin{figure}[]
    \centering
    \includegraphics[width=\linewidth]{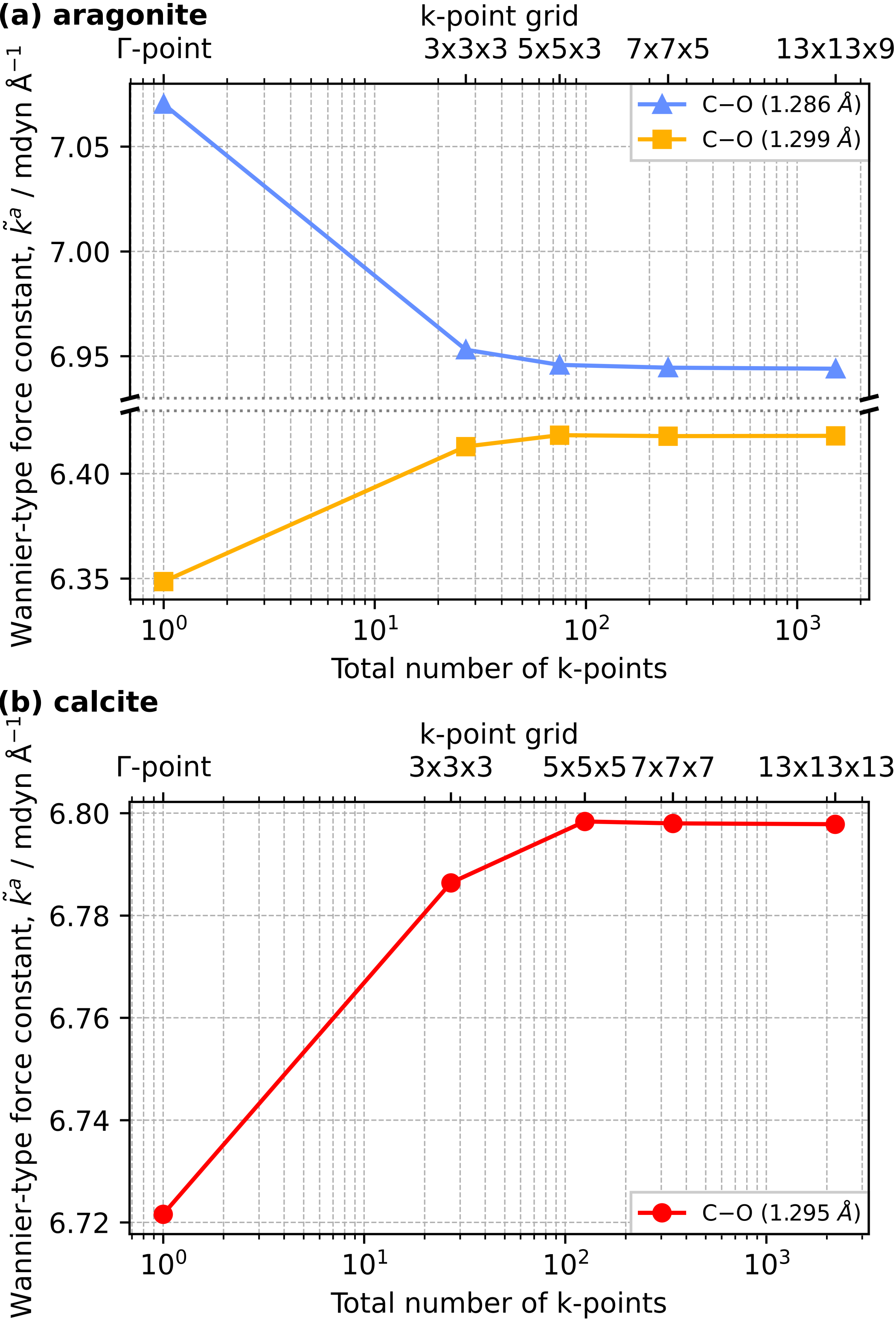}
    \begin{subfigure}{0pt}
        \phantomcaption
        \label{fig:caco3-CO:arag}
    \end{subfigure}
    \begin{subfigure}{0pt}
        \phantomcaption
        \label{fig:caco3-CO:calc}
    \end{subfigure}
    \caption{Convergence of Wannier-type force constants for the \ce{C-O} bonds in aragonite and calcite when computed on increasingly dense reducible Brillouin-zone grids.}
    \label{fig:caco3-CO}
\end{figure}

The \ce{CO3^2-} anions sit on crystallographic sites of site symmetry $m$ in the aragonite structure, breaking the symmetry of the \ce{C-O} bonds into two distinct subgroups, \Cref{fig:caco3-CO:arag}. For the first group, the $\Gamma$-point $k^a(\mathbf{k}=\Gamma)$ are notably higher than the converged $\tilde{k}^a=6.944$~mdyn~\AA$^{-1}$ obtained from a dense sampling of the RBZ. The second group of \ce{C-O} bonds exhibit the opposite behavior, with $k^a(\mathbf{k}=\Gamma)$ underestimating the converged $\tilde{k}^a=6.418$~mdyn~\AA$^{-1}$. In the calcite structure the \ce{CO3^2-} anions sit on three-fold symmetric sites (site symmetry $3$), conserving the symmetry of the \ce{C-O} bonds, with $k^a(\mathbf{k}=\Gamma)$ underestimating the converged $\tilde{k}^a = 6.798$~mdyn~\AA$^{-1}$, \Cref{fig:caco3-CO:calc}. Notably, our results show that including the contributions from the full phonon dispersion in the local mode construction can have significant effects on the predicted bond strengths that are unpredictable \emph{a priori}.

We can further compare the values of $\tilde{k}^a$ obtained for the crystalline polymorphs to that obtained for the gas-phase anion, $k^a \approx 6.511$~mdyn~\AA$^{-1}$ (see ESI S3.7), noting that $\tilde{k}^a$ does not exist for gas-phase structures. This comparison indicates that crystallization in fact changes the effective stiffness of the \ce{C-O} bonds of the anion in both polymorphs. Importantly, the magnitude of the crystallization effect on the \ce{C-O} bond stiffness would be poorly captured if $k^a(\mathbf{k}=\Gamma)$ were used for the analysis of the crystalline phase. We therefore find that vectors $\tilde{k}^a$ do provide a tool to quantitatively resolve chemically meaningful variations in bond strength both within a single crystal structure and between polymorphic forms. 

\begin{figure}[]
    \centering
    \includegraphics[width=\linewidth]{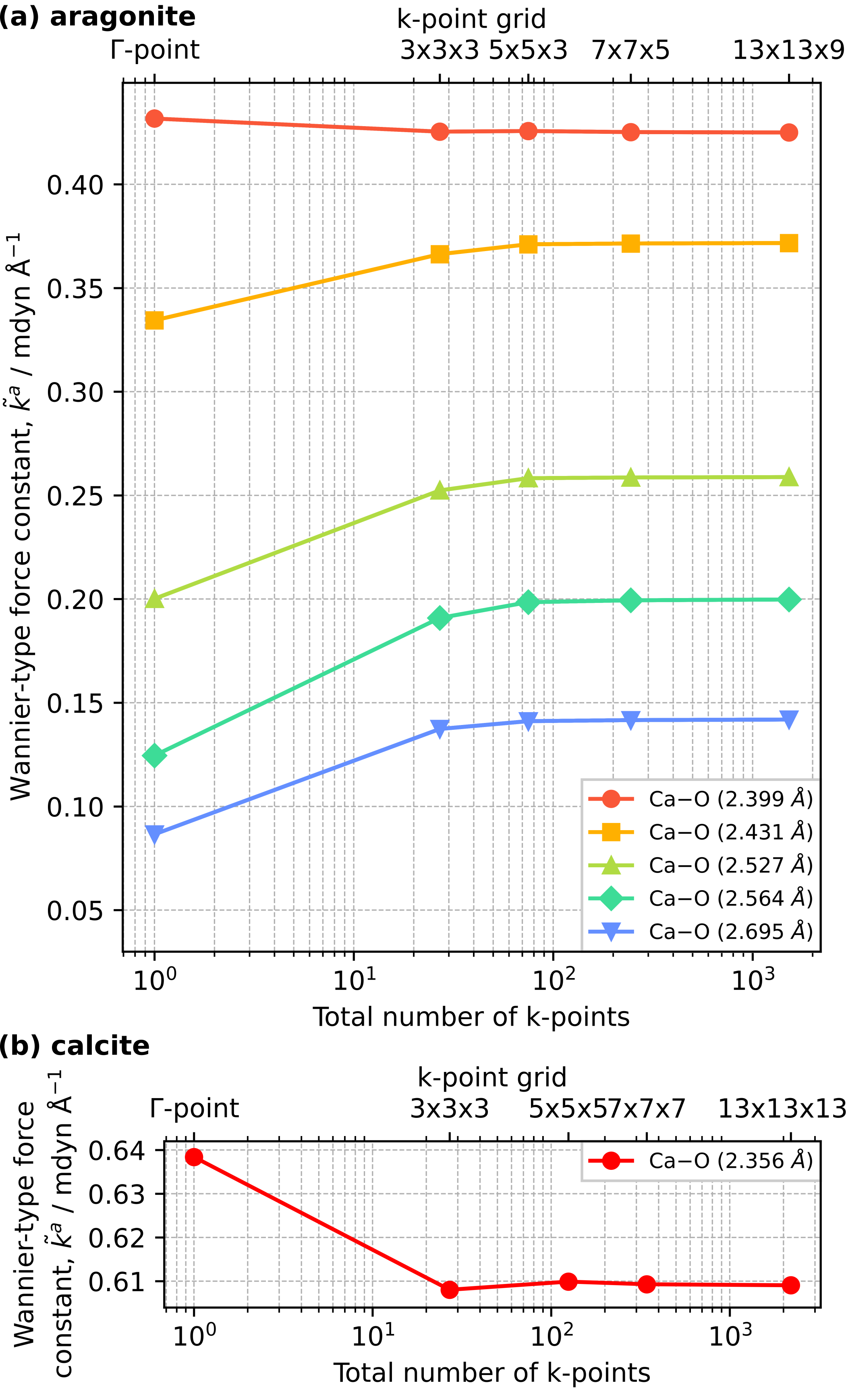}
    \begin{subfigure}{0pt}
        \phantomcaption
        \label{fig:caco3-CaO:arag}
    \end{subfigure}
    \begin{subfigure}{0pt}
        \phantomcaption
        \label{fig:caco3-CaO:calc}
    \end{subfigure}
    \caption{Convergence of Wannier-type force constants for the \ce{Ca-O} bonds in aragonite and calcite when computed on increasingly dense reducible Brillouin-zone grids.}
    \label{fig:caco3-CaO}
\end{figure}

We can also compare the \ce{Ca^2+}$\cdots$\ce{O^2-} interactions in the two polymorphic forms. In aragonite, the nine-coordinate \ce{Ca^2+} coordination sits on a crystallographic site with symmetry $m$, leading to five unique \ce{Ca^2+}$\cdots$\ce{O^2-} interactions for each cation: four sets of degenerate pairs across the mirror plane and one unique bond in the mirror plane, \Cref{fig:caco3-CaO:arag}. For the degenerate pairs, analysis of \hbox{$k^a(\mathbf{k}=\Gamma)$} underestimates the converged values of $\tilde{k}^a$ (which vary from $0.142$--$0.425$~mdyn~\AA$^{-1}$), albeit to different degrees. The \ce{Ca^2+} octahedra in calcite instead sit on sites of $\bar3$ symmetry, ensuring all six \ce{Ca^2+}$\cdots$\ce{O^2-} interactions are equivalent. In this case $k^a(\mathbf{k}=\Gamma)$ slightly overestimate the stiffness of the \ce{Ca^2+}$\cdots$\ce{O^2-} interactions as compared with the converged value of $\tilde{k}^a = 0.609$~mdyn~\AA$^{-1}$, \Cref{fig:caco3-CaO:calc}. Our analysis thus demonstrates that $\tilde{k}^a$ offers a useful and intuitive approach to quantify the effects of local coordination environment on bonding strengths in solids.

Overall, these results demonstrate that Wannier-type local mode force constants are highly sensitive to the local chemical environment and coordination chemistry. Whilst a construction of $k^a(\mathbf{k}=\Gamma)$ by sampling only the $\Gamma$-point does provide some insight into the bonding behavior in solids, the magnitude of the predicted values can be unreliable. This is particularly true for periodic systems that exhibit high degrees of phonon dispersion, such as inorganic and extended solids. Since the Wannier-type local mode force constants provide a closer physical picture of bond strength in solids as compared with our earlier wavevector-resolved approach,\cite{mojsak_jctc_2026} we therefore encourage the use of the former where bond strength information is the target.

\section{Conclusions}

Building upon our recent development of wavevector-resolved local vibrational mode theory, we have presented here an approach to spatially localize internal-coordinate vibrations in periodic systems using a Wannier-type procedure. Together, these two representations provide complementary tools to analyze lattice dynamics: while wavevector‑resolved local modes elucidate how specific internal coordinates participate in collective phonon dispersion throughout the Brillouin zone, the Wannier‑type construction yields real‑space localized vibrational modes centered on a single unit cell. In this way, the presented Wannier-type construction provides a more direct and chemically intuitive approach for quantifying the strength of specific bonds and interactions in periodic systems, consistent with the interpretation of LVMT in isolated molecules.

The original construction of wavevector-resolved local modes provides a fortuitous approach to rectifying the gauge problem of conventional Wannier approaches, and makes our analysis readily applicable across chemically and structurally diverse periodic systems. By demonstrating our method on a range of representative crystalline systems, we have shown that local mode force constants obtained from analysis of only the $\Gamma$-point normal modes can vary significantly from the more physically-intuitive local mode force constants obtained from the Wannier-type construction. The magnitude and sign of this variability is linked to the degree of phonon dispersion across the crystal Brillouin zone, and thus cannot, in general, be reliably estimated \textit{a priori}. We do, however, note that for the studied systems, the qualitative trends in relative interaction strengths predicted using both the $\Gamma$-point-only and Wannier-type approaches are conserved. This suggests that $\Gamma$-point local mode analyses may still provide useful qualitative insight when full Brillouin-zone sampling is computationally prohibitive.

At present, the method is formulated within the harmonic approximation and requires access to phonon dispersions across the Brillouin zone. As such, the computational cost of accessing Wannier-type local mode force constants as bond strength indicators depends heavily on the complexity of the system of interest and the method used to perform lattice dynamics calculations. However, the Wannier‑type construction itself is trivial to perform at virtually no additional cost where simulated phonon dispersion relations are already available. 

Overall, the combined use of wavevector‑resolved and Wannier‑type local vibrational mode analyses offers a rich and comprehensive framework for understanding structure-property relations in crystalline solids. The present Wannier-type formulation establishes a direct real‑space analog of molecular local vibrational modes for periodic materials, enabling chemically intuitive modeling of bond and interaction strengths from the response of the system to localized internal-coordinate distortions. In this respect we expect this framework to become a broadly applicable tool to connect crystal structure, lattice dynamics, and chemical bonding in periodic materials.

\begin{acknowledgments}
The computations described in this paper were performed using the University of Birmingham's BlueBEAR HPC service, which provides a High Performance Computing service to the University's research community. See www.birmingham.ac.uk/bear for more details. Additional calculations were made possible via the UK Materials and Molecular Modelling Hub, which is partially funded by the EPSRC (EP/T022213/1). MM thanks the University of Birmingham for the award of a PhD studentship. AALM thanks BAM for a Wilhelm-Ostwald Fellowship. This material is based upon work supported by the Air Force Office of Scientific Research under award number FA8655-25-1-7065.
\end{acknowledgments}

~\\
\section*{Supplemental Material}
Raw data are available in the Zenodo repository at: link. The post-processing script implementing the Wannier-type construction from wavevector-resolved local modes is available at our group GitHub, link. Further mathematical derivations and simulation details are available in the electronic Supplemental Information document.

\bibliography{references}

\end{document}